\documentstyle[epsfig]{article}
\def\be{\begin{equation}}
\def\ee{\end{equation}}
\def\ba{\begin{eqnarray}}
\def\ea{\end{eqnarray}}

\newcommand{\eq}{\begin{equation}}
\newcommand{\en}{\end{equation}}
\newcommand{\eqa}{\begin{eqnarray}}
\newcommand{\ena}{\end{eqnarray}}
\newcommand{\bea}{\begin{eqnarray}}
\newcommand{\eea}{\end{eqnarray}}

\newcommand{\ZZ}{\hbox{{\rm Z{\hbox to 3pt{\hss\rm Z}}}}}
\newcommand{\NP}[1]{Nucl.\ Phys.\ {\bf #1}}

\newcommand{\PR}[1]{Phys.\ Rev.\ {\bf #1}}

\begin{document}
\begin{titlepage}
\begin{flushright}
DFTT 7/97\\
January 1997
\end{flushright}
\vskip0.5cm
\begin{center}
{\Large\bf  The Svetitsky--Yaffe conjecture}\\ 
{\Large\bf for the plaquette operator.}
\end{center}
\vskip 0.6cm
\centerline{F.Gliozzi and P.Provero}
\vskip 0.6cm
\centerline{\sl Dipartimento di Fisica
Teorica dell'Universit\`a di Torino}
\centerline{\sl Istituto Nazionale di Fisica Nucleare, Sezione di Torino}
\centerline{\sl via P.Giuria 1, I--10125 Torino, Italy
\footnote{e--mail:gliozzi, provero~@to.infn.it}}
\vskip 0.6cm
\begin{abstract}
According to the Svetitsky--Yaffe conjecture, a $(d+1)$--dimensional 
pure gauge theory undergoing a continuous deconfinement transition is in 
the same universality class as a $d$--dimensional statistical model 
with order parameter taking values in the center of the gauge group.
We show that the plaquette operator of the gauge theory is mapped into
the energy operator of the statistical model. For $d=2$, this 
identification allows us to use conformal field theory techniques to 
evaluate exactly the correlation functions of the plaquette operator 
at the critical point. In particular, we can evaluate exactly the
plaquette expectation value in presence of static sources, which 
gives some new insight in the structure of the color flux tube in
mesons and baryons.
\end{abstract}
\end{titlepage}
\setcounter{footnote}{0}
\def\thefootnote{\arabic{footnote}}
\section{Introduction}
Consider a $(d+1)$--dimensional pure gauge theory undergoing a continuous 
deconfinement transition at the critical temperature $T_{c}$. The 
effective model describing the behavior of Polyakov lines at finite 
temperature $T$ will be a $d$--dimensional statistical model with 
global symmetry group coinciding with the center of the gauge group. 
Svetitsky and Yaffe \cite{sy} were able to show that this effective 
model has only short--range interactions. 
If also the $d$--dimensional effective model displays 
a continuous phase transition then it follows
from universality arguments 
that it belongs to the same universality 
class of the original gauge model. 
\par
Therefore all the universal properties of the deconfinement 
transition can be predicted to coincide with the ones of the 
dimensionally reduced effective model. These include the values of the 
critical indices, the finite--size scaling behavior, and the 
correlation functions at criticality. 
The conjecture has passed several numerical tests, which became more 
and more stringent in the last years due to the increased precision 
reachable with Monte Carlo simulations (see \cite{ch} and 
references therein). 
\par
It is clear that the Svetitsky--Yaffe conjecture becomes very 
predictive for $d=2$, where, using the methods of conformal field 
theory, the critical behavior can be determined exactly. For example, 
the critical properties of $(2+1)D$ $SU(2)$ gauge theory 
at the deconfinement temperature coincide with those of the 
two--dimensional Ising model. This allows us not only to predict the 
exact values of the critical indices, but also to write down all the 
multipoint correlation functions of the Polyakov loop at criticality 
\cite{nando}. 
\par
What is needed to fully exploit the predictive power 
of the Svetitsky--Yaffe conjecture is a mapping relating 
the physical observables of the 
gauge theory to the operators of the dimensionally reduced 
model. In $d=2$ the knowledge of this mapping is equivalent to 
solving the gauge theory at the deconfinement temperature.
\par
The correspondence between Polyakov line and order parameter of the 
effective model is the first entry in this mapping and is 
intrinsically contained in the Svetitsky--Yaffe conjecture. It is 
natural to ask what operator in the $d$--dimensional model 
corresponds to the plaquette operator of the gauge theory: symmetry 
considerations suggest the energy operator as a natural candidate. In 
this paper we show that this is actually the case, and we describe 
some consequences of this identification.
\par
The correctness of the identification plaquette--energy is shown in Sec. 
2 by studying the finite--size behavior of the plaquette operator in 
$(2+1)D$ $Z_{2}$ gauge theory at the critical temperature. We show that it 
coincides with the (highly non--trivial) finite-size behavior of the 
energy operator in the $2D$ Ising model at criticality. 
\par
In Sec. 3 we compute correlation functions of the plaquette operator 
by using conformal field theory techniques. In particular, the 
expectation value of the plaquette in the vacuum modified by static 
sources can be computed for $(2+1)D$ $SU(2)$ and $SU(3)$ gauge theories 
at the deconfinement temperature, providing physical insight about the 
structure of the color flux--tube in mesons and baryons.
\section{Finite--size behavior of the plaquette expectation value}
Finite--size effects at criticality are typically rather strong, due 
to scale invariance, and non--trivial. 
Therefore they are ideally suited to compare theoretical predictions 
with, for example, results of Monte Carlo simulations.
 In particular, for two--dimensional statistical systems,
the critical behavior, including finite--size effects, is completely 
understood with the methods of conformal field theory (CFT). We want 
to exploit this fact to establish the correspondence between the 
plaquette operator in a $(d+1)$--dimensional lattice gauge theory at 
the deconfinement transition and the energy operator of the 
corresponding $d$--dimensional statistical model.
\par
Consider for example the $2D$ Ising model: 
the shape and size dependence  at criticality
of the expectation value of the internal
energy on a torus is given by \cite{ff,dif,id}
\begin{equation}
\langle \epsilon\rangle = 
\frac{\pi\sqrt{\Im m \tau}\left|\eta(\tau)\right|^{2}}
{\sqrt{A}Z_{1/2}(\tau)}\label{fse}
\end{equation}
where $A$ and $\tau$ are respectively the area and the modular 
parameter of the torus, and $Z_{1/2}$ is the Ising partition function 
at the critical point:
\begin{equation}
Z_{1/2}=\frac{1}{2}\sum_{\nu=2}^{4}\left|\frac{\theta_{\nu}(0,\tau)}
{\eta(\tau)}\right|
\end{equation}
where $\theta_{\nu}$ are the Jacobi theta functions and $\eta$ is the
Dedekind function (for notations and conventions see Ref.\cite{id}).
\par
Comparing Eq.(\ref{fse}) with the finite--size behavior of the 
plaquette operator in a $3D$ lattice gauge theory such that the 
center of the gauge group is $Z_{2}$ provides a stringent test of our 
identification. The simplest choice is the $3D$ $Z_{2}$ gauge model, 
for which it is possible to achieve very high precision in the Monte 
Carlo evaluation of physical quantities, and accurate estimates of the 
deconfinement temperature are available. 
\par
Therefore we considered the $(2+1)D$ $Z_{2}$ gauge model on lattices of 
size $L_{1}\times L_{2}\times L_{t}$ where $L_{t}\ll L_{1},L_{2}$
with periodic boundary conditions on all directions and 
we studied it at the critical coupling $\beta_{c}(L_{t})$, which is 
known to high accuracy for several values of $L_{t}$ \cite{ch}. By 
performing Monte Carlo simulations at different values of 
$L_{1},L_{2}$, we can compare the finite--size behavior of the 
plaquette expectation value with Eq.(\ref{fse}).
\par
More precisely, we will show that the plaquette operator is a 
mixture of the identity and energy operators of the $2D$ CFT: on one 
hand, both these operators transform as singlets under $Z_{2}$, and 
therefore can contribute to the plaquette operator; on the other 
hand, we know that the plaquette expectation value does not vanish in 
infinite volume, unlike the energy operator of the $2D$ Ising model 
(see Eq.(\ref{fse})). Therefore, a non--vanishing contribution of the 
identity operator must be expected in the plaquette expectation value.
Hence our conjecture is
\be
\langle \Box \rangle =c_{1} \langle 1\rangle
+c_{\epsilon}\langle\epsilon\rangle
\ee
where the expectation value in the l.h.s. is taken in the LGT, while 
the ones in the r.h.s. refer to the CFT.
The prediction for the finite--size behavior of the plaquette 
expectation value is therefore
\be
\langle \Box \rangle_{L_{1}L_{2}}=c_{1}+c_{\epsilon}\frac{F(\tau)}
{\sqrt{L_{1}L_{2}}}+O(1/L_1 L_2)\label{fseplaq}
\ee
where $F$ is a function of the modular parameter only:
\be
F(\tau)=\frac{\pi \sqrt{\Im m\tau}\left|\eta(\tau)\right|^2}
{Z_{1/2}(\tau)}
\ee
\par 
We performed our Monte Carlo simulations at $L_{t}=6$
 with critical coupling
\be
\beta_c (L_t=6)=0.746035
\ee
\par
We measured the plaquette expectation value for lattices of
area $100\le A \le 6400$ and asymmetry ratio $\Im m \tau=1,2,4$.
The space--like and time--like plaquettes have different expectation
values, therefore must be fitted separately with Eq.(\ref{fseplaq}).
The Monte Carlo simulations were actually performed in the $3D$ Ising
(spin) model, which is exactly equivalent through duality  
to the $Z_2$ gauge model.
This choice allowed us to use a non--local cluster simulation algorithm.
\par
The agreement is very 
good, giving $\chi^{2}_{red}=0.7$ for space--like plaquettes and
$\chi^{2}_{red}=0.9$ for time--like plaquettes. 
\footnote{
It must be noted that 
the expectation values of space--like and time--like
plaquettes for a given lattice are not 
statistically uncorrelated, since they were extracted form the same sample of
configurations.
}
These data are plotted in Fig. 1.
\par
\begin{figure}[p]
\begin{center}
\mbox{\epsfig{file=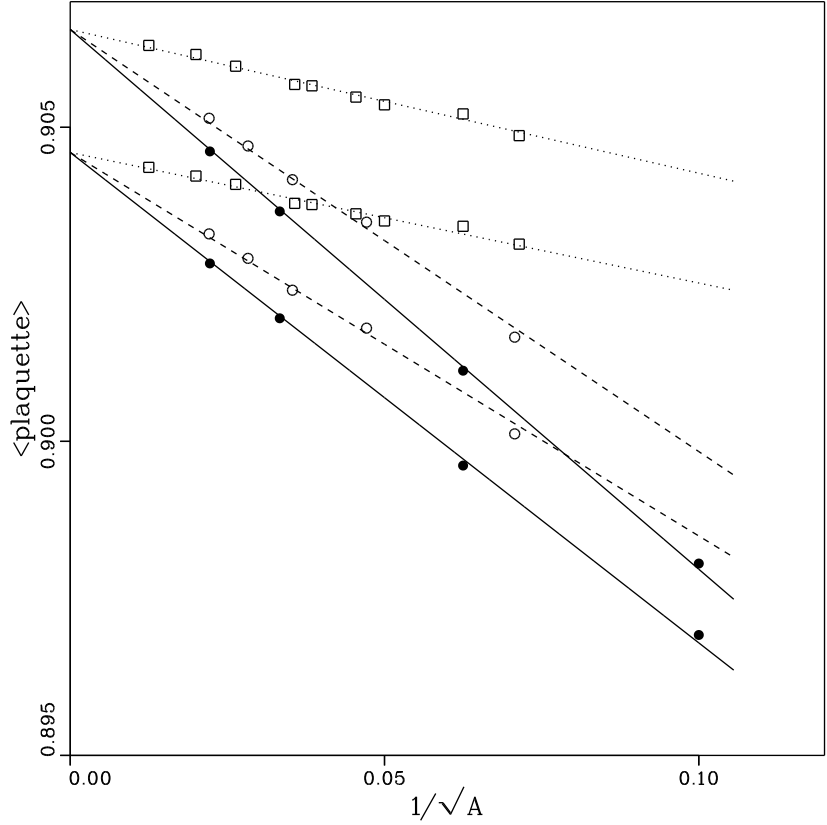}}
\vskip 2mm
\caption
{\it 
Size and shape dependence of the plaquette expectation value. Black dots
correspond to square lattices ($\Im m\tau=1$). White dots and squares 
correspond to rectangular lattices with $\Im m\tau=2$ and $\Im m \tau=4$
respectively. Both time--like and space--like plaquettes are shown,
the latter having lower expectation values. The lines correspond to the
best fit to Eq.(\ref{fseplaq}).
}
\end{center}
\end{figure}
\section{Correlation functions of the plaquette operator}
In this section we will exploit the new entry we added to the Svetitsky--Yaffe 
mapping to compute correlation functions of the plaquette operator at 
the deconfinement temperature. This will provide some new insight into 
the structure of color flux tubes in mesons and baryons.
\par
Consider for example $(2+1)D$ $SU(2)$ LGT at the deconfinement temperature. To 
study the flux tube structure in a ``static meson'' we can consider 
the plaquette expectation value in the vacuum modified by the 
presence of two static sources, {\em i. e.} the correlation function 
of the plaquette operator with two Polyakov loops:
\be
G(x,x_{1},x_{2})=\langle\Box(x) P(x_{1})P(x_{2}) \rangle-
\langle \Box\rangle\langle P(x_{1})P(x_{2}) \rangle
\ee
where $x,x_{1},x_{2}$ are points in the $2D$ space. 
This will be given by the correlation of the energy operator with 2 
spin operators in the $2D$ critical Ising model:
\be
G(x,x_{1},x_{2})\propto \langle \epsilon(x) \sigma(x_{1})\sigma(x_{2})
\rangle_{Ising}
\ee
The r.h.s. is easily computed in CFT and we find
\be
G(x,x_{1},x_{2})\propto\frac{\left|x_{1}-x_{2}\right|^{3/8}}
{\left(\left|x-x_{1}\right|\left|x-x_{2}\right|\right)^{1/2}}
\label{meson}
\ee
We have plotted this function in Fig. 2.
\par
\begin{figure}[htb]
\begin{center}
\mbox{\epsfig{file=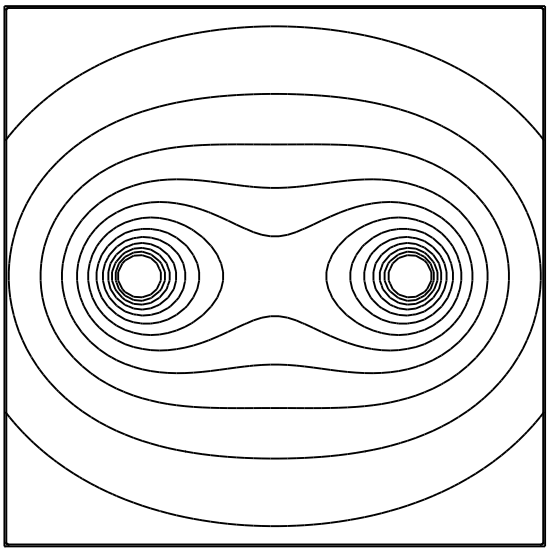}}
\vskip 2mm
\caption
{\it 
Structure of the flux tube in a "static meson" at the deconfinement
temperature, Eq.(\ref{meson}).
}
\end{center}
\end{figure}
More interesting is the case of $(2+1)D$ $SU(3)$ LGT, where we can 
consider a ``static baryon'' by modifying the vacuum with three 
static sources and compute
\be
G(x,x_{1},x_{2},x_{3})=\langle\Box(x) P(x_{1})P(x_{2})P(x_{3}) \rangle
-\langle\Box \rangle\langle P(x_{1})P(x_{2})P(x_{3}) \rangle
\ee
Our identification gives
\be
G(x,x_{1},x_{2},x_{3})\propto \langle \epsilon(x) \sigma(x_{1})\sigma(x_{2})
\sigma(x_{3})\rangle_{3-state\  Potts}
\ee
where the correlation function on the r.h.s. must be computed in the 
$c=4/5$ CFT describing the three-state Potts model at criticality. 
This is done using the methods introduced in Ref.\cite{dots} (see 
also \cite{id}) and gives
\ba
G(x,x_{1},x_{2},x_{3})&\propto&\frac{\left(\left|x_{1}-x_{2}\right|
\left|x_{1}-x_{3}\right|\left|x_{2}-x_{3}\right|\right)^{1/15}}
{\left(\left|x-x_{1}\right|\left|x-x_{2}\right|\left|x-x_{3}\right|
\right)^{4/15}}\nonumber\\
&&\left|y(1-y)\right|^{7/15}
\left[|f_{1}(y)|^2+
\frac{9}{4}\frac{\Gamma^{3}(3/5)\Gamma(1/5)}
{\Gamma^{3}(2/5)\Gamma(4/5)}\left|f_{2}(y)\right|^2\right]\label{baryon} 
\ea
where, introducing a complex coordinate $z$ in $2D$ space,
$y$ is the conformally invariant cross-ratio
\be
y=\frac{(z-z_{1})(z_{2}-z_{3})}{(z-z_{3})(z_{2}-z_{1})}
\ee
and $f_{1}$ and $f_{2}$ are hypergeometric functions:
\ba 
f_{1}(y)&=&F(4/5,7/5;8/5;y)\\
f_{2}(y)&=&y^{-3/5}F(1/5,4/5;2/5;y)
\ea
\par
$G(x,x_{1},x_{2},x_{3})$ is plotted in Fig. 3, for the case in which the three 
static sources form an equilateral triangle. Notice that this 
calculation brings strong support to the ``Y'' structure of flux tube 
in baryons (see {\em e.g.} \cite{y} and references therein), 
as opposed to the ``$\Delta$'' structure 
\cite{delta}. 
\par
\begin{figure}[htb]
\begin{center}
\mbox{\epsfig{file=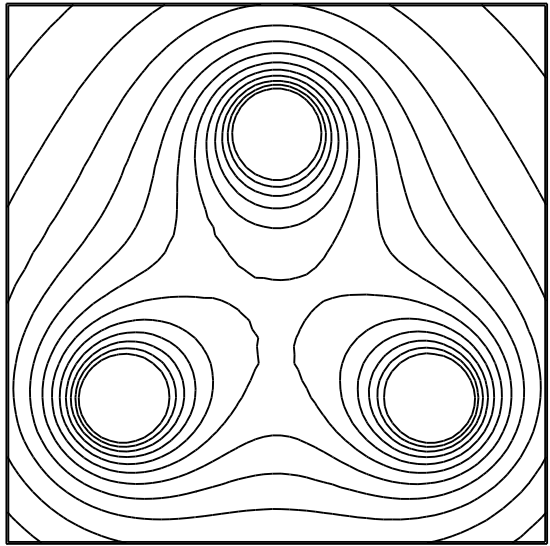}}
\vskip 2mm
\caption
{\it The structure of the flux tube in a "static baryon", 
Eq. (\ref{baryon}).}
\end{center}
\end{figure}
\section{Conclusions}
In this paper we have added a new entry to the Svetitsky--Yaffe 
mapping between $(d+1)$--dimensional LGT's at the deconfinement 
temperature and $d$--dimensional critical statistical models, namely 
we have shown that the plaquette operator of the LGT is mapped into 
the energy operator of the statistical model. 
\par
For $d=2$, this identification  allows  in principle the exact 
evaluation of all  correlations of the plaquette operator at the 
deconfinement point, providing a useful tool for the study of the 
color flux tube in mesons and baryons.
\vskip1.cm
We would like to thank M. Caselle and M. Hasenbusch for useful discussions. 
This work has been supported in part by the European Commission TMR 
programme ERBFMRX-CT96-0045
and by the Ministero ita\-lia\-no 
del\-l'Uni\-ver\-si\-t\`a e della Ricerca Scientifica e Tecnologica.


\begin{thebibliography}{99}
\bibitem{sy}B.Svetitsky and L.G.Yaffe, \NP{B210} (1982) 423.
\bibitem{ch}M.Caselle and M.Hasenbusch, \NP{B470} (1996) 435.
\bibitem{nando}F.Gliozzi and S.Vinti, contribution to Lattice 96, 
hep-lat/9609026.
\bibitem{ff}A.E.Ferdinand and H.G.Fisher, \PR{185} (1969) 832.
\bibitem{dif}P.Di Francesco, H.Saleur and J.B.Zuber, \NP{B290} (1987)
527.
\bibitem{id}C.Itzykson and J.Drouffe, ``Statistical Field Theory'',
Cambridge 1989, Chap. 9.
\bibitem{dots}V.S. Dotsenko, \NP{B235} (1983) 54.
\bibitem{y} Yu.S.Kalashnikova and A.V.Nefediev, hep-ph/9604411.
\bibitem{delta} J.M.Cornwall, \PR{D54} (1996) 6527.
\end{thebibliography}
\end{document}